\documentclass[twocolumn,showpacs,preprintnumbers,amsmath,amssymb,aps]{revtex4}
\usepackage{graphicx}

\begin{document}

\title{Direct measurement of quantum phases in graphene via photoemission spectroscopy}
\author{Choongyu Hwang,$^{1,4}$}
\author{Cheol-Hwan Park,$^{2,4}$}
\author{David A. Siegel,$^{1,2}$}
\author{Alexei V. Fedorov,$^3$}
\author{Steven G. Louie$^{1,2}$} \email{sglouie@berkeley.edu}
\author{Alessandra Lanzara$^{1,2}$} \email{ALanzara@lbl.gov}

\affiliation{$^1$Materials Sciences Division, Lawrence Berkeley
National Laboratory, Berkeley, California 94720, USA.}
\affiliation{$^2$Department of Physics, University of California at
Berkeley, Berkeley, California 94720, USA.}
\affiliation{$^3$Advanced Light Source, Lawrence Berkeley National
Laboratory, Berkeley, California 94720, USA.}
\affiliation{$^4$These authors contributed equally to this work.}

\date{\today}

\begin{abstract}
Quantum phases provide us with important information for
understanding the fundamental properties of a system. However, the
observation of quantum phases, such as Berry's phase and the sign of
the matrix element of the Hamiltonian between two non-equivalent
localized orbitals in a tight-binding formalism, has been challenged
by the presence of other factors, e.\,g.\,, dynamic phases and
spin/valley degeneracy, and the absence of methodology. Here, we
report a new way to directly access these quantum phases, through
polarization-dependent angle-resolved photoemission spectroscopy
(ARPES), using graphene as a prototypical two-dimensional material.
We show that the momentum- and polarization-dependent spectral
intensity provides direct measurements of (i) the phase of the band
wavefunction and (ii) the sign of matrix elements for non-equivalent
orbitals. Upon rotating light polarization by $\pi/2$, we found that
graphene with a Berry's phase of $n\,\pi$ ($n=$~1 for single- and
$n=$~2 for double-layer graphene for Bloch wavefunction in the
commonly used form) exhibits the rotation of ARPES intensity by
$\pi/n$, and that ARPES signals reveal the signs of the matrix
elements in both single- and double-layer graphene. The method
provides a new technique to directly extract fundamental quantum
electronic information on a variety of materials.
\end{abstract}


\maketitle

\section{Introduction}

Quantum phases are the most beautiful example of quantum physics and
essential to understand physics in any material. For example,
Berry's phase, the accumulated phase in the eigenfunction acquired
by evolving the quantum system adiabatically around a closed loop in
the parameter space of the Hamiltonian~\cite{Bohm}, has been shown
to be responsible for the Aharonov-Bohm effect~\cite{Aha}, the
half-integer quantum Hall effect~\cite{Zhang,Novoselov,McCann}, etc.
Another important example is the sign of the hopping matrix element
(or hopping integral) $\left< \phi_1\right|H\left|\phi_2\right>$ of
the Hamiltonian between two non-equivalent localized orbitals
$\phi_1$ and $\phi_2$ in a tight-binding formalism. This phase, a
fundamental quantity in determining the electronic structure of a
system, is dictated by the characteristics of the atomistic
interaction, e.\,g.\,, whether it is attractive or repulsive. Both
of these are important to directly extract fundamental quantum
electronic information on a variety of
materials~\cite{Taguchi,Murakami,Hsieh}.

In graphene, the Berry's phase is theoretically extracted from the
spinor eigenstate, which are $\pi$ for single- and $2\pi$ for
double-layer graphene~\cite{McCann}. From recent studies, the $n\pi$
Berry's phase in the commonly used form of the spinor states is, in
fact, related to the pseudospin winding number $n$ of a particle as
it travels in a loop in $k$-space which encloses the Dirac
point~\cite{Parkcond}. These values have been measured through
magneto-transport experiments~\cite{Zhang,Novoselov} that are
typically neither capable of measuring Berry's phase of a specific
electron bandstructure nor free from spin/valley degeneracy of the
electron bandstructure of a system under study. Additionally, this
method requires a strong magnetic field, which breaks time-reversal
symmetry in graphene. Meanwhile, the signs of hopping integrals
between non-equivalent orbitals for graphene (graphite) have only
been determined by {\it ab initio} calculations~\cite{Gruneis},
e.\,g.\,, using maximally localized Wannier
functions~\cite{Marzari}. Since the sign of hopping integral depends
on the characteristics of the localized orbitals and the interaction
between them, it is crucial in determining the electron
bandstructure within a tight-binding formalism. However, the absence
of methodology has led to use the signs following the well-known
Slonczewski-Weiss-McClure model~\cite{SWMc1,SWMc2} without
experimental verification for the past few decades. Additionally,
the sign of hopping integral between non-equivalent orbitals has
never been determined experimentally for any material.

Given the high momentum-resolving power of angle-resolved
photoemission spectroscopy (ARPES), ARPES is an ideal candidate to
solve the above issues on the determination of quantum phases. For
example, the phase difference between the matrix elements describing
two different optical transitions at the (110) surface of platinum
was extracted from a combined study of a spin-resolved ARPES
experiment and a theoretical model~\cite{YuPt}. Also, ARPES has been
employed to study the characteristics of the spinor eigenstates in
graphite~\cite{Himpsel} and graphene~\cite{Eli}, which revealed an
interference effect between photo-excited
electrons~\cite{Himpsel,Eli}. However, these theoretical studies,
within a tight-binding formalism, have incorrectly treated the
interaction Hamiltonian, which is the key part in the photoemission
process as it describes the interaction between photons and
electrons. Moreover, it has not been clear how Berry's phase enters
in ARPES intensity and the sign of hopping integral has only been
speculated without any comparison with experiments~\cite{Eli}, which
naturally leads to incorrect values.

Here we report that ARPES can indeed provide information on these
quantum phases, e.\,g.\,, the Berry's phase and the sign of hopping
integral between non-equivalent orbitals. The phase factor in the
spinor eigenstate of graphene~\cite{McCann} gives rise to strong
intensity variation around a constant energy contour. Upon rotating
light polarization by $\pi/2$, we found that graphene with a Berry's
phase of $n\,\pi$ ($n=$~1 for single- and $n=$~2 for double-layer
graphene) exhibits the rotation of ARPES intensity maxima by
$\pi/n$, which gives important advantages compared to the
conventional magneto-transport method~\cite{Zhang,Novoselov}.
Additionally, we found that full polarization-dependence of ARPES
signal reveals the sign of hopping integrals in both single- and
double-layer graphene (graphite can also be understood), e.\,g.\,,
$\gamma'_0>0$ and $\gamma'_1>0$, which is the first experimental
determination of the sign of hopping integral between non-equivalent
orbitals for any material by any method.

\section{Sample preparation}
Single- and double-layer graphene samples were grown epitaxially on
$n$-doped 6$H$-SiC(0001) surfaces by electron-beam heating, as
detailed elsewhere~\cite{Rollings}. An SiC sample was mounted in a
prep-chamber with a base pressure of 5$\times$10$^{-10}$ Torr to
remove a thick oxide layer from the sample by heating at 600
$^{\circ}$C for a few hours. The clean sample was then transferred
to a custom-designed chamber equipped with low-energy-electron
microscopy (LEEM) with a base pressure of 2$\times$10$^{-11}$ Torr
and heated to 1000 $^{\circ}$C under Si flux to improve the surface
conditions for graphene growth. The sample temperature was raised to
1400 $^{\circ}$C or 1600 $^{\circ}$C (determined by an optical
pyrometer) to make single- or double-layer graphene, respectively.

  \begin{figure}[b]
  \begin{center}
  \includegraphics[width=0.9\columnwidth]{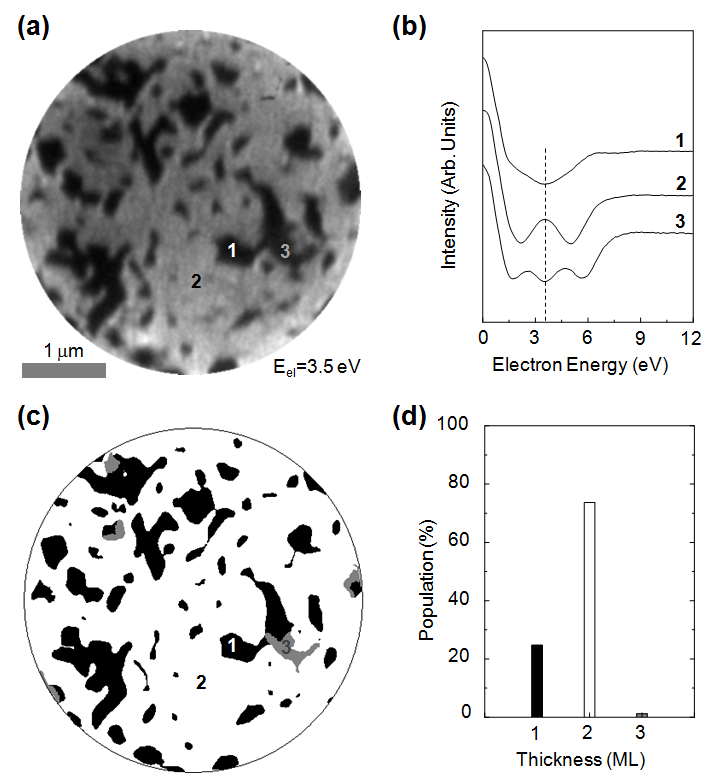}
  \end{center}
  \caption{
(a) A LEEM image using an electron energy of 3.5~eV over
4$\mu$m$\times$4$\mu$m range. (b) Reflectivity spectra for the three
regions (1, 2, and~3) specified in (a). (c) Post-processed image of
(a) showing the regions covered by single-, double-, and
triple-layer graphene. (d) A histogram showing the fractions of
single-, double-, and triple-layer graphene in our sample used for
double-layer graphene measurements.}
  \label{Fig1}
  \end{figure}

The surface morphology was monitored {\it in situ} during the sample
growth by LEEM at the National Center for Electron Microscopy at
Lawrence Berkeley National Laboratory. The thickness of fabricated
graphene samples was determined by LEEM measurements performed at
room temperature following the standard
procedure~\cite{David,Ohta2}. In particular, the electron
reflectivity versus kinetic energy curve varies significantly with
the number of graphene layers providing position-dependent
measurements on the number of graphene layers. A typical bright
field image for double layer graphene is shown in Fig.~\ref{Fig1}(a)
over 4 $\mu$m$\times$4 $\mu$m range, recorded with electron beam of
kinetic energy 3.5~eV denoted as the dashed line in
Fig.~\ref{Fig1}(b). In order to determine the number of graphene
layers at each position, the electron reflectivity is plotted as a
function of electron kinetic energy, as shown in Fig.~\ref{Fig1}(b),
where the number of dips is the same as the number of graphene
layers. Regions~1, 2, and~3 in Fig.~1(a) show 1, 2, and 3 dips,
respectively, corresponding to single-, double-, or triple-layer
graphene, respectively. These regions are painted in black, white,
and grey, respectively, in Fig.~\ref{Fig1}(c). The fractions of
regions in the sample covered by different numbers of graphene
layers were determined from the areal fractions of differently
colored regions in Fig.~\ref{Fig1}(c).  In particular, we find that
the double-layer graphene sample contains $\sim$74~\% of
double-layer and $\sim$22~\% of single-layer graphene.

%%%%%%%%%%%%%%%%%%%%%%%%%%%%%%%%%%%%%%%%%%%%%%%%%%%%%%%%%%%%%%%%%%%
  \begin{figure*}
  \begin{center}
  \includegraphics[width=1.8\columnwidth]{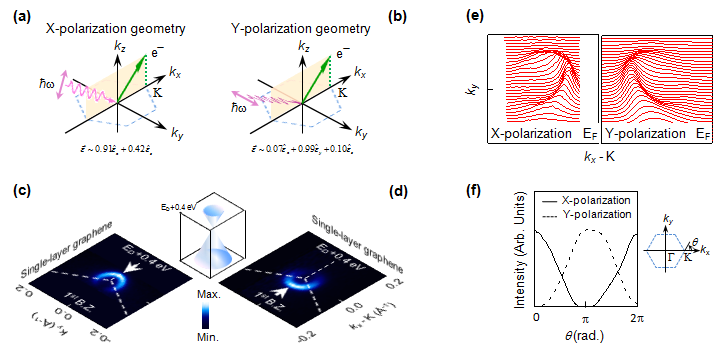}
  \end{center}
  \caption{
(a) X-polarization geometry. (b) Y-polarization geometry. A beam of
monochromatic lights with energy $\hbar\omega=$50~eV and
polarization vector $\vec{\varepsilon}$ is incident on a sample. The
light polarizations in X- and Y-polarization geometries are almost
parallel to the {\it x} and {\it y} axes, respectively. (c, d)
Measured intensity maps of single-layer graphene at energy $E=E_{\rm
F}$ with X- and Y-polarized lights, respectively. Intensity maxima
are denoted by white arrows and the electronic band structure of
single-layer graphene is drawn in the cartoon. (e) Constant-energy
ARPES intensity maps for single-layer graphene at $E_{\rm F}$ with
X- and Y-polarized light. (f) The angle-dependent intensity profiles
of single-layer graphene are obtained by integrating the
constant-energy intensity map along the radial direction from the
Dirac point, in which the solid and dashed lines denote the
experimental data for X- and Y-polarized lights, respectively. The
angle $\theta$ is measured from the $+k_x$ direction. The plotted
quantities are with respect to the intensity minimum.}
  \label{Fig2}
  \end{figure*}
%%%%%%%%%%%%%%%%%%%%%%%%%%%%%%%%%%%%%%%%%%%%%%%%%%%%%%%%%%%%%%%%%

\section{EXPERIMENT}

We have performed polarization-dependent ARPES experiments on
single- and double-layer graphene at 10 K using a photon energy of
50 eV at beam-lines 10.0.1 and 12.0.1 of Advanced Light Source at
Lawrence Berkeley National Laboratory. In Figs.~2(a) and~2(b), we
show the typical geometry of ARPES experiments: a beam of
monochromatized light with energy $\hbar\omega$ and polarization
vector $\vec{\varepsilon}$ is incident on a sample, resulting in the
emission of photoelectrons in all directions. The polarization
vector of light is referenced with respect to the sample normal. In
the experiment presented here, two different geometries were
employed as shown in Figs.~\ref{Fig2}(a) and~\ref{Fig2}(b). In one
geometry shown in Fig.~\ref{Fig2}(a), the polarization of light is
almost parallel to the {\it x} axis, while in the other shown in
Fig.~\ref{Fig2}(b) to the {\it y} axis; hence, we define these two
geometries as X- and Y-polarization, respectively. These geometries
have the advantage with respect to the conventional {\it s}- and
{\it p}-polarizations used in previous studies
~\cite{Pescia,Prince}, to measure the whole two-dimensional
variation of the intensity maps around a singular (Dirac) point and
not just the intensity distributions along two characteristic lines
in momentum space. This aspect becomes particularly relevant for
some experimental conditions, e.\,g.\,, photon energy and sample
orientation (i.\,e.\,, the mixture of light polarizations), when the
intensity maps (or initial electronic states) are neither symmetric
nor anti-symmetric with respect to the reflection plane.  Under this
condition in fact, the conventional notations would not give
appropriate information on the symmetry of the initial states.

Figures~\ref{Fig2}(c) and~\ref{Fig2}(d) show the experimental
photoelectron intensity maps at the Fermi level, $E_{\rm F}$, versus
the two-dimensional wavevector {\bf k} for single-layer graphene,
for the two polarization geometries. Here, $E_{\rm F}$ is 0.4~eV
above the Dirac point energy, $E_{\rm D}$~\cite{Zhou,David,Seyller}.
The main feature in the intensity maps of both geometries is an
almost circular Fermi surface centered at the K point as shown in
Figs.~\ref{Fig2}(c) and~\ref{Fig2}(d), as expected for a conical
dispersion. This is in good agreement with a recent
polarization-dependent ARPES study on single-layer graphene when
using photons with energy lower than 52 eV~\cite{Gierz}.
Surprisingly we find that the the angular intensity distribution is
quite different for the two polarizations: for the X-polarization
geometry, the minimum intensity position is in the first Brillouin
zone, whereas for the Y-polarization geometry, the maximum intensity
position is in the first Brillouin zone, suggesting a $\pi$ rotation
of the maximum intensity in the $k_x$-$k_y$ plane around the K point
upon rotating the light polarization by $\pi/2$, from X to Y (see
the white arrows).

The exact rotation angle is extracted from the direct comparison
between the raw momentum distribution curves (MDCs) in
Fig.~\ref{Fig2}(e) and the angular dependence of the photoelectron
intensity maps integrated over the radial direction around the {\bf
K} point in Fig.~\ref{Fig2}(f). There is an overall shift of the
intensity maxima (minima) by $\sim\pi$ upon changing the light
polarization from X (black solid line) to Y (black dashed line),
although the latter appears to be slightly shifted by $\sim\pi/10$
with respect to $\pi$. As we will show later, this is due to the
presence of a finite polarization component along the $k_x$
direction in our experimental geometry.

%%%%%%%%%%%%%%%%%%%%%%%%%%%%%%%%%%%%%%%%%%%%%%%%%%%%%%%%%%%%%%%
  \begin{figure*}
  \begin{center}
  \includegraphics[width=1.8\columnwidth]{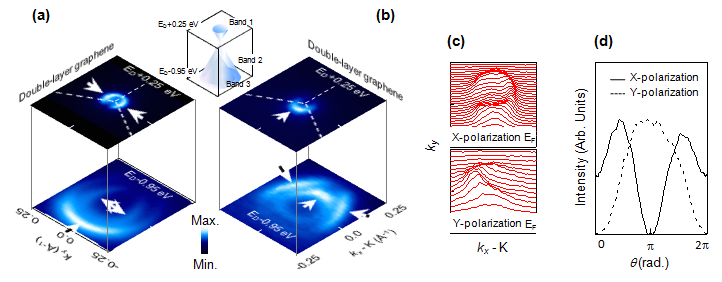}
  \end{center}
  \caption{
(a, b) Measured intensity maps of double-layer graphene at energies
$E=E_{\rm D}+0.25$~eV$~(=E_{\rm F})$ and $E=E_{\rm D}-0.95$~eV.
Intensity maxima are denoted by white arrows and the electronic band
structure of double-layer graphene is drawn in the cartoon. (c)
Measured intensity maps of double-layer graphene at energy $E=E_{\rm
F}$. (d) The angle-dependent intensity profiles of double-layer
graphene at energy $E=E_{\rm F}$, in which the solid and dashed
lines denote the experimental data for X- and Y-polarized lights,
respectively. The plotted quantities are with respect to the
intensity minimum.}
  \label{Fig3}
  \end{figure*}
%%%%%%%%%%%%%%%%%%%%%%%%%%%%%%%%%%%%%%%%%%%%%%%%%%%%%%%%%%%%%%%%%

We note that not only the angular position of the intensity maximum,
but also the absolute value of it changes upon changing light
polarization. The maximum intensity ratio from experiments is
X-polarized/Y-polarized=21.4. However, this number itself is not
very meaningful, because the measured ARPES intensity is affected by
the difference in the experimental geometries for X- and Y-polarized
lights (the difference in the out-of-plane component of light
polarization, photon flux per unit area, etc., which are the factors
that cannot be controlled to be the same in different experimental
geometries). On the other hand, the ratio from our theory that will
be discussed later provides X-polarized/Y-polarized=0.83, assuming
that the experimental parameters for two geometries are the same
except for the in-plane light polarization.

A similar study on Bernal stacked double-layer graphene reveals a
strong and complicated momentum-, band-, and polarization-dependence
as shown in Figs.~\ref{Fig3}(a) and~\ref{Fig3}(b), that is
qualitatively different from that of single-layer graphene. Like
single-layer graphene, the double-layer sample is slightly $n$
doped~\cite{Seyller}, therefore, only three of the four $\pi$ bands
are occupied and hence detectable with ARPES as shown in the cartoon
of Fig.~\ref{Fig3}(a). The most prominent feature is that, when the
light polarization is changed from X to Y, the maximum intensity
positions around the K point in the $k_x$-$k_y$ plane are rotated by
$\sim\pi/2$ (see white arrows at $E_{\rm F}$). This is in striking
contrast with the single-layer case where the rotation is $\sim\pi$
as seen from the raw MDCs in Fig.~\ref{Fig3}(c) and the
photoelectron intensity maps integrated over the radial direction
around the {\bf K} point shown in Fig.~\ref{Fig3}(d). Due to
trigonal warping effects~\cite{Antonio}, however, the rotation for
higher-energy states is not exactly $\pi/2$ as shown in
Figs.~\ref{Fig3}(a) and~\ref{Fig3}(b).

\section{Theoretical analysis}

\begin{figure}[t]
  \begin{center}
  \includegraphics[width=1.0\columnwidth]{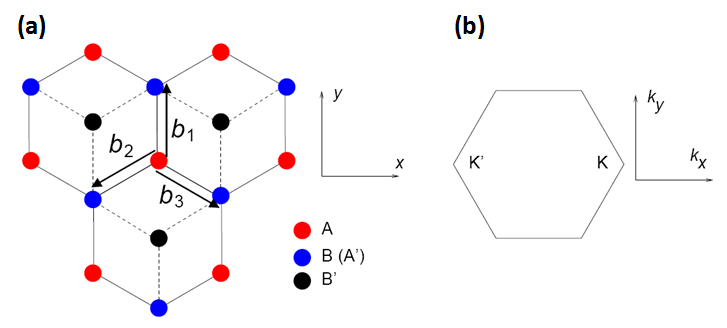}
  \end{center}
  \caption{
(a) Schematic of single- and double-layer graphene. (b) The
Brillouin zone. Here, ${\bf b}_1=b\left(0,1\right)$, ${\bf
b}_2=b\left(-\frac{\sqrt{3}}{2},-\frac{1}{2}\right)$
    and ${\bf b}_3=b\left(\frac{\sqrt{3}}{2},-\frac{1}{2}\right)$ are the three vectors
    connecting the in-plane nearest neighbor atoms
    where the inter-carbon distance $b=1.42$~\AA, and the lattice constant is $a=\sqrt{3}b$.
    The positions of the K and K$'$ points are
    $\left(\frac{4\pi}{3a},0\right)$ and
    $\left(-\frac{4\pi}{3a},0\right)$, respectively.}
  \label{FigS1}
  \end{figure}

To the best of our knowledge, the only models in the literature
describing the polarization-dependence of the ARPES intensity in
graphite \cite{Himpsel} and graphene \cite{Eli} are substantially
different from our results. Previous studies, in fact, predict a
small polarization-dependence~\cite{Himpsel} and no
polarization-dependence~\cite{Eli} of the photoelectron intensity
maps, respectively. Therefore, to be able to reproduce our
experimental findings and understand what lies behind this
nontrivial polarization dependence, we have developed a new model.
In particular, we first consider the Hamiltonian using the
tight-binding model based on the $p_{\rm z}$ orbital of each carbon
atom using two parameters: $t_0$ and $t_1$ for the in-plane
nearest-neighbor (A-B or A$'$-B$'$) and the inter-layer vertical
(B-A$'$) hopping integrals, respectively, as schematically drawn in
Fig.~\ref{FigS1}(a). The parameters $t_0$ and $t_1$ correspond to
$-\gamma_0'$ and $\gamma_1'$, respectively, in the well-known
Slonczewski-Weiss-McClure (SWMc) model~\cite{SWMc1,SWMc2}. In our
calculation, we have used $|t_0|=3.16$~eV and $|t_1|=0.39$~eV, which
are the values in Table II of Gr\"uneis {\it et al}.~\cite{Gruneis},
but we do not fix the signs of them. Note that all four possible
choices of the signs give exactly the same electron energy band
structure within this two-parameter tight-binding model.

%%%%%%%%%%%%%%%%%%%%%%%%%%%%%%%%%%%%%%%%%%%%%%%%%%%%%%%%%%%%%%%%%%%%%%%%%%
%%%%%%%%%%%%%%%%%%%%%%%%%%%%%%%%%%%%%%%%%%%%%%%%%%%%%%%%%%%%%%%%%%%%%%%5

With this setup, the tight-binding Hamiltonian of a double-layer
graphene for two-dimensional wave vector ${\bf k}=(k_{\rm x},k_{\rm
y})$ using a basis set composed of Bloch sums of localized orbitals
on each of the four sublattices (A, B, A$'$, and B$'$) reads
\begin{equation}
H^0_{\rm double}({\bf k})=\left(
\begin{array}{cccc}
0 & t_0\,g({\bf k}) & 0 & 0\\
\\
t_0\,g^*({\bf k}) & 0 & t_1\,e^{-i\,k_{\rm
z}\,d} & 0\\
\\
0 & t_1\,e^{i\,k_{\rm
z}\,d} & 0 & t_0\,g({\bf k})\\
\\
0 & 0 & t_0\,g^*({\bf k}) & 0\\
\end{array}
\right)\,. \label{eq:H_bi_0}
\end{equation}
Here,
\begin{equation}
g({\bf k})=\sum_{i=1}^{3}\exp(i{\bf k}\cdot{\bf b}_i)
\label{eq:g_def}
\end{equation}
with ${\bf b}_i$'s defined as in Fig.~\ref{FigS1}(a), and
\begin{equation}
\left(
\begin{array}{c}
1\\
0\\
0\\
0
\end{array}
\right)_{\bf k} =\frac{1}{\sqrt{N}}\sum_{{\bf R}_{\rm A}} e^{i{\bf
k}\cdot{\bf R}_{\rm A}}\phi({\bf r}-{\bf R}_{\rm A})\,,
\label{eq:1000}
\end{equation}
\begin{equation}
\left(
\begin{array}{c}
0\\
1\\
0\\
0
\end{array}
\right)_{\bf k} =\frac{1}{\sqrt{N}}\sum_{{\bf R}_{\rm B}} e^{i{\bf
k}\cdot{\bf R}_{\rm B}}\phi({\bf r}-{\bf R}_{\rm B})\,,
\label{eq:0100}
\end{equation}
etc. We note that often the {\bf k} dependence of the basis function
is suppressed in the spinor notation for simplicity. In
Eq.~(\ref{eq:H_bi_0}), we have considered a phase difference $e^{\pm
i\,k_{\rm z}\,d}$ arising from the finite inter-layer distance {\it
d} and the perpendicular component of electron wave vector $k_{\rm
z}$. Here, $k_{\rm z}$ is not part of the Bloch momentum, but the
$z$ component of the photoelectron wave vector, which is determined
by the photon energy. This quantity plays a crucial role in
determining the ARPES intensity of double-layer graphene as will be
discussed later and also of multi-layer graphene as previously
reported~\cite{OhtaPRL}.

The additional interaction Hamiltonian coupling to electromagnetic
waves of wavevector {\bf Q} for a double-layer graphene
$\hat{H}_{\rm double}^{\rm int}$ can be obtained by
%taking the first-order terms of {\bf A} from
%$H^0_{\rm bi}\left({\bf k}-\frac{e}{\hbar c}{\bf A}\right)-H^0_{\rm bi}({\bf k})$.
using the velocity operator $\hat{\bf v}=\left[\hat{\bf
r},\hat{H}_{\rm double}^0\right]/i\hbar$, where $\hat{\bf
r}=i\hbar\,(\nabla_{\bf k},\partial_{k_{\rm z}})$ in the {\bf
k}-representation and $\hbar$ is the Planck's constant, as
$-\frac{e}{c}\hat{\bf A}\cdot\hat{\bf v}$~\cite{Yu} [{\it e} is the
charge of an electron, {\it c} is the speed of light, and the
external vector potential is given by ${\bf A}({\bf r},t)={\bf
A_Q}e^{i({\bf Q}\cdot{\bf r}-\omega t)}$]~\cite{Louie} , i.\,e.\,,
\begin{widetext}
\begin{equation}
H_{\rm double}^{\rm int}({\bf k},{\bf Q})=-\frac{e}{\hbar\,c} {\bf
A}_{\bf Q}\cdot \left(
\begin{array}{cccc}
0 & t_0\,\nabla_{\bf k}g({\bf k}) & 0 & 0\\
\\
t_0\,\nabla_{\bf k}g^*({\bf k}) & 0 & -i\,d\,t_1\,e^{-ik_{\rm
z}d}\,\hat{z} & 0\\
\\
0 & i\,d\,t_1\,e^{ik_{\rm
z}d}\,\hat{z} & 0 & t_0\,\nabla_{\bf k}g({\bf k})\\
\\
0 & 0 & t_0\,\nabla_{\bf k}g^*({\bf k}) & 0\\
\end{array}
\right)\,. \label{eq:H_bi_int}
\end{equation}
\end{widetext}
The transition matrix elements in Eq.~(\ref{eq:H_bi_int}) are those
taken between basis functions of Bloch sums of $p_{\rm z}$ orbitals
of wavevectors ${\bf k}+{\bf Q}$ and {\bf k}.
Equation~(\ref{eq:H_bi_int}) is valid when $1/|{\bf Q}|$ is much
larger than the distance between the nearest-neighbor atoms $b$,
i.\,e.\,, when the variation in ${\bf A}({\bf r},t)$ over a length
scale of {\it b} is much smaller than ${\bf A}({\bf r},t)$ itself.
We shall eventually take the ${\bf Q}\to0$ limit in our discussion
because the momentum of light is negligible for photon energies
considered.

For single-layer graphene, performing a similar type of analysis, we
obtain
\begin{equation}
H^0_{\rm single}({\bf k})=\left(
\begin{array}{cc}
0 & t_0 g({\bf k})\\
\\
t_0\,g^*({\bf k}) & 0\\
\end{array}
\right)\,, \label{eq:H_mono_0}
\end{equation}
and
\begin{equation}
H_{\rm single}^{\rm int}({\bf k},{\bf Q})=-\frac{e}{\hbar\,c} {\bf
A_Q}\cdot \left(
\begin{array}{cccc}
0 & t_0\,\nabla_{\bf k}g({\bf k})\\
\\
t_0\,\nabla_{\bf k}g^*({\bf k}) & 0\\
\end{array}
\right)\,. \label{eq:H_mono_int}
\end{equation}

\begin{figure*}
  \begin{center}
  \includegraphics[width=1.8\columnwidth]{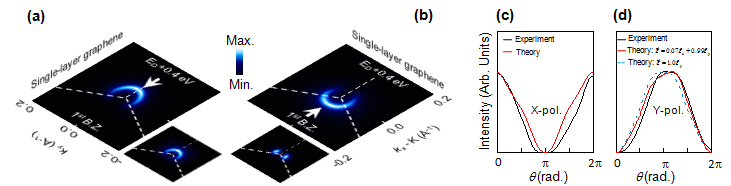}
  \end{center}
  \caption{
(a, b) Calculated intensity maps of single-layer graphene for X- and
Y-polarized lights, respectively. The insets are the results of
calculations~\cite{Himpsel} using the simplified momentum operator
instead of the correct velocity operator. An arbitrary energy
broadening of 0.10~eV has been used. Intensity maxima are denoted by
white arrows. (c, d) The angle-dependent intensity map of
single-layer graphene for X- and Y-polarized lights, respectively,
in which the solid black, solid red, and dashed blue lines denote
the experimental data, the calculated results obtained by assuming
the actual light polarization used in the experiment, and the
calculated results obtained by assuming perfectly Y-polarized light,
respectively. The theory results shown in (a) and (b) have adopted
the light polarization used in the actual experiment shown in
Figs.~\ref{Fig2}(a) and~\ref{Fig2}(b)). The plotted quantities are
with respect to the intensity minimum.}
  \label{Fig4}
  \end{figure*}

\begin{figure*}[ht]
  \begin{center}
  \includegraphics[width=1.8\columnwidth]{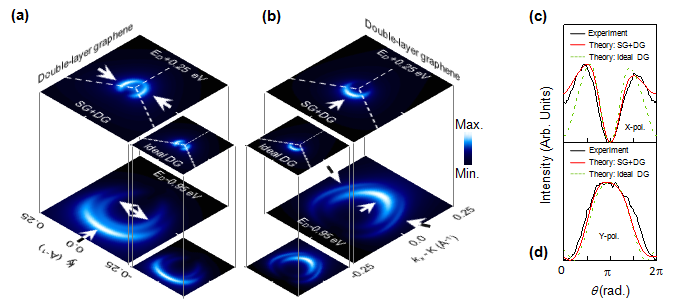}
  \end{center}
  \caption{
(a, b) Calculated intensity maps of double-layer graphene for X- and
Y-polarized lights, respectively. An arbitrary energy broadening of
0.10, 0.10, and 0.15~eV have been used for bands~1, 2, and~3,
respectively. Panels denoted by `Ideal DG' show results obtained by
considering a double-layer graphene alone and those denoted by
'SG$+$DG' show results considering the contribution from some
single-layer fraction of the sample as well (see text). Intensity
maxima are denoted by white arrows. (c, d) The angle-dependent
intensity map of single-layer graphene for X- and Y-polarized
lights, respectively, in which the solid black, solid red, and
dashed green lines denote the experimental data, the calculated
results obtained by considering the contribution from single-layer
graphene portion of the sample, and the calculated results for ideal
double-layer graphene, respectively. The theory results shown in (a)
and (b) have adopted the light polarization used in the actual
experiment as shown in Figs.~\ref{Fig2}(a) and~\ref{Fig2}(b). The
plotted quantities are with respect to the intensity minimum.}
  \label{Fig5}
  \end{figure*}

$H^{\rm int}$ is critical to explain the polarization dependence of
$I_{{\bf k}}$, because it describes the interaction between
electrons and photons. The lack of polarization dependence in
previous studies \cite{Himpsel,Eli} is indeed due to the way in
which $H^{\rm int}$ is incorrectly treated. In one case~\cite{Eli},
the light interaction is completely neglected by setting $H^{\rm
int}=1$, while in the earlier study on graphite~\cite{Himpsel}, the
velocity operator ${\bf v}$ is replaced by the momentum ${\bf
p}/m_0=-i\hbar\nabla/m_0$, where $\hbar$ is the Planck's constant
and $m_0$ the free-electron mass. This replacement
works~\cite{Starace,Louie} only when the Hamiltonian is local,
whereas a tight-binding Hamiltonian, which has been used in the
previous studies~\cite{Himpsel,Eli} as well as our study, is
intrinsically non local. The experimental finding of a strong
polarization dependence of $I_{{\bf k}}$ in Figs.~\ref{Fig2}
and~\ref{Fig3} clearly shows the need for a more complete
theoretical treatment. We have developed a theory using the widely
adopted tight binding model with one $p_z$-like localized orbital
per carbon atom, but employing the appropriate interaction
Hamiltonian with the velocity operator. A very good agreement
between our model and the experimental results is obtained for all
polarizations and for both single- and double-layer graphene, when
we compare Fig.~\ref{Fig2} with Fig.~\ref{Fig4} and Fig.~\ref{Fig3}
with Fig.~\ref{Fig5} as will be discussed later.

In order to understand what lies behind the observed non-trivial and
unexpected wavevector-dependent photoelectron intensity $I_{{\bf
k}}$, we need to calculate the absolute square of the transition
matrix element $M_{s\,{\bf k}}=\left<f_{{\bf k}+{\bf
Q}}\right|H^{\rm int}({\bf k},{\bf Q}) \left|\psi_{s\,{\bf
k}}\right>$, where $\left|\psi_{s\,{\bf k}}\right>$ is a single- or
double-layer graphene eigenstate with $s=\pm1$ the band index,
$\left| f_{{\bf k}+{\bf Q}}\right>$ is the plane-wave final state
projected onto the $p_{\rm z}$ orbitals of graphene [both
$\left|\psi_{s\,{\bf k}}\right>$ and $\left| f_{{\bf k}+{\bf
Q}}\right>$ are expressed using the basis set of Bloch sums of
localized $p_z$ orbitals at sublattices A and B in
Fig.~\ref{Fig4}(a)] and $H^{\rm int}=-\frac{e}{c}{\bf A}\cdot{\bf
v}$~\cite{Yu}, which should not be neglected in photoemission
process~\cite{Eli}. The use of a projection of the final plane-wave
state onto the Bloch sum, which -- when using plane-waves basis --
is effectively composed of multiple plane-waves~\cite{CH}, allow to
explain the non-trivial polarization dependence of the ARPES
intensity distribution in Figs.~2(c) and~2(d).  Since the
polarization of {\bf A} is in the {\it x-y} plane, the projection of
$\left|f_{\bf k}\right>$ onto the $\sigma$-states of graphene will
result in zero contribution to the transition matrix elements and
hence are neglected in this analysis. For simplicity of notation,
and without any loss of generality, in the rest of this section, we
shall take the limit of ${\bf Q}\to0$ and denote $H^{\rm int}({\bf
k})=H^{\rm int}({\bf k},{\bf Q})$ and $\left| f_{{\bf
k}}\right>=\left| f_{{\bf k}+{\bf Q}}\right>$. For single-layer
graphene, we may use
\begin{equation}
\left| f_{\bf k}\right> = \frac{1}{\sqrt{2}} \left(
\begin{array}{c}
1\\
1
\end{array}
\right)_{\bf k} \label{eq:pw_mono}
\end{equation}
and for double-layer graphene
\begin{equation}
\left| f_{\bf k}\right> = \frac{1}{2} \left(
\begin{array}{c}
1\\
1\\
1\\
1
\end{array}
\right)_{\bf k}\,. \label{eq:pw_bi}
\end{equation}
For photons (with energy $\approx50$~eV) used in the experiment,
$k_{\rm z}$ of the planewave final state is much larger than the
variation of two-dimensional Bloch wavevector {\bf k} around a
single Dirac point, leading to only a small variation in $k_{\rm z}$
with any change in {\bf k} around a single Dirac point. For light
with a nonzero polarization component along the {\it z} direction,
it will only give rise to an additive isotropic term to the
photoelectron intensity that is independent of the in-plane
polarization of the light.

%%%%%%%%%%%%%%%%%%%%%%%%%%%%%%%%%%%%%%%%%%%%%%%%%%%%%%%%%%%%%%%%%
%%%%%%%%%%%%%%%%%%%%%%%%%%%%%%%%%%%%%%%%%%%%%%%%%%%%%%%%%%%%%%%%%

Now, let us consider the case where {\bf k} is very close to the
Dirac point K as shown in Fig.~\ref{FigS1}(b), and define ${\bf
q}={\bf k}-{\bf K}$ ($|{\bf q}|\ll |{\bf K}|$). According to
Eq.~(\ref{eq:g_def}),
\begin{equation}
g({\bf q}+{\bf K})\approx
-\frac{\sqrt{3}}{2}a\left(q_x-iq_y\right)\,, \label{eq:g}
\end{equation}
where {\it a} is the lattice parameter. For single-layer graphene,
therefore,
\begin{equation}
H^0_{\rm single}({\bf q}+{\bf K})\approx -\frac{\sqrt{3}}{2}a t_0
\left( q_x\,\sigma_x+q_y\,\sigma_y \right)\,, \label{eq:H_mono_0_K}
\end{equation}
and
\begin{equation}
H_{\rm single}^{\rm int}({\bf q}+{\bf K})\approx
\frac{\sqrt{3}e}{2\hbar c}a t_0 \left(
A_{0\,x}\,\sigma_x+A_{0\,y}\,\sigma_y \right)\,,
\label{eq:H_mono_int_K}
\end{equation}
where $\sigma_x$ and $\sigma_y$ are the Pauli matrices. The energy
eigenvalue and wavefunction of Eq.~(\ref{eq:H_mono_0_K}) are given
by $E_{s\,{\bf k}}=\frac{\sqrt{3}}{2}a |t_0|\,s\, |{\bf q}|$ and
\begin{equation}
\left|\psi_{s\,{\bf k}}\right>=\frac{1}{\sqrt{2}} \left(
\begin{array}{c}
e^{-i\theta_{\bf q}/2}
\\
-{\rm sgn}(t_0)\,s\,e^{i\theta_{\bf q}/2}
\end{array}
\right)\,, \label{eq:wfn_mono}
\end{equation}
respectively, when $\theta_{\bf q}$ is the angle between {\bf q} and
the $+k_{\rm x}$ direction. Using Eqs.~(\ref{eq:pw_mono}),
(\ref{eq:H_mono_int_K}), and~(\ref{eq:wfn_mono}), the transition
matrix element is given for light polarized along the {\it x}
direction by
\begin{equation}
M^{x-{\rm pol}}_{s\,{\bf k}}\sim \exp({-i\theta_{\bf q}/2}) -{\rm
sgn}(t_0)\,s\,\exp(i\theta_{\bf q}/2)\,. \label{eq:mtx_mono}
\end{equation}
It follows that for $s=+1$ (states above the Dirac point energy),
\begin{equation}
|M^{x-{\rm pol}}_{+1\,{\bf k}}|^2\sim\sin^2(\theta_{\bf q}/2)
\label{eq:mtx2_mono_1}
\end{equation}
and
\begin{equation}
|M^{x-{\rm pol}}_{+1\,{\bf k}}|^2\sim\cos^2(\theta_{\bf q}/2)
\label{eq:mtx2_mono_2}
\end{equation}
with $t_0>0$ and $t_0<0$, respectively.

Similarly, for light polarized along the {\it y} direction, the
transition matrix element is given by
\begin{equation}
M^{y-{\rm pol}}_{s\,{\bf k}}\sim \exp(-i\theta_{\bf q}/2) +{\rm
sgn}(t_0)\,s\,\exp(i\theta_{\bf q}/2)\,. \label{eq:mty_mono}
\end{equation}
It follows that for $s=+1$ (states above the Dirac point energy),
\begin{equation}
|M^{y-{\rm pol}}_{+1\,{\bf k}}|^2\sim\cos^2(\theta_{\bf q}/2)
\label{eq:mty2_mono_1}
\end{equation}
and
\begin{equation}
|M^{y-{\rm pol}}_{+1\,{\bf k}}|^2\sim\sin^2(\theta_{\bf q}/2)
\label{eq:mty2_mono_2}
\end{equation}
with $t_0>0$ and $t_0<0$, respectively.

In both cases, we can explain the rotation of the intensity maximum
in the photoelectron intensity map around the K point by $\pi$ upon
the change from X- to Y-polarized light. Comparing
Eqs.~(\ref{eq:mtx2_mono_1}) and~(\ref{eq:mtx2_mono_2}) with
Eqs.~(\ref{eq:mty2_mono_1}) and~(\ref{eq:mty2_mono_2}), irrespective
of the sign of $t_0$, the maxima of the photoemission intensity map
of single-layer graphene is rotated by $\pi$ when the light
polarization is rotated by $\pi/2$, in agreement with experiment
shown in Fig.~\ref{Fig2}. Moreover, the theoretical results with
$t_0<0$ shown in Figs.~\ref{Fig4}(a) and~\ref{Fig4}(b) agree with
the measured angular spectral intensity shown in Figs.~\ref{Fig2}(c)
and~\ref{Fig2}(d); especially, the choice of $|t_0|=3.16$~eV (fitted
to previous experiments~\cite{Gruneis}) reproduces quite well the
salient features in the experimentally measured intensity maps.

This is even more clear from the angular dependence of theoretical
photoelectron intensity drawn with the red solid lines compared to
experimental results drawn with the black solid lines in
Figs.~\ref{Fig4}(c) and~\ref{Fig4}(d) . Note that experimental
intensity maximum for Y-polarized light shows additional shift by
$\sim\pi/10$ in Fig.~\ref{Fig2}(f). This additional shift is well
understood by a finite polarization component along the $k_x$
direction. When we assume the actual light polarization used in the
experiment shown in Fig.~\ref{Fig2}(b), the theoretical intensity
exactly matches with the experimental result. On the other hand,
when we assume the ideal Y-polarization, the intensity maximum
appears at $\pi$ as shown with the blue dashed line in in
Fig.~\ref{Fig4}(d). Therefore, we determine through experiment that
the inter-orbital hopping matrix element $t_0$ between two in-plane
nearest-neighbor carbon $p_{\rm z}$ orbitals is negative; we will
come back to this point later.

Recent theoretical study on the matrix element in single-layer
graphene~\cite{Gierz} has found that, in order to describe the
matrix element for Y-polarized light from first-principles
calculations using plane-wave basis, one needs multiple plane-wave
components for the final photo-emitted electron state. Since we
consider a projection of the final state onto the Bloch sum, which
-- when using the plane-waves basis -- is effectively composed of
multiple plane-waves~\cite{CH}, our approach can explain the
non-trivial polarization dependence of the ARPES intensity
distribution in Figs.~2(c) and~2(d). Additionally, our result
obtained by using photons with energy 50 eV is in good agreement
with the recent study~\cite{Gierz} based on first-principles
calculations using photons with energy lower than 52 eV. This
suggests that, at photon energy below 52 eV and only when the
correct interaction Hamiltonian is employed, the projection of the
final state onto the tight-binding Bloch sum may describe the true
final state qualitatively. However, since our theoretical framework
is within tight-binding formulation, and not from first-principles
calculations with plane-wave basis, convergence tests with respect
to the number of plane-waves and the character of the true final
state are beyond the scope of this paper and, in fact, has been done
in a recent study~\cite{Gierz}.

%%%%%%%%%%%%%%%%%%%%%%%%%%%%%%%%%%%%%%%%%%%%%%%%%%%%%%%%%%%%%%%%
%%%%%%%%%%%%%%%%%%%%%%%%%%%%%%%%%%%%%%%%%%%%%%%%%%%%%%%%%%%%%%%%

In the case of double-layer graphene, for simplicity of the
analysis, we confine our discussion to the inner parabolic bands
(bands~1 and~2 in the cartoon in Fig.~\ref{Fig3}), although we
considered all the four bands in our theoretical calculations shown
in Figs.~\ref{Fig5}(a) and~\ref{Fig5}(b). In double-layer graphene,
for electronic states with energy $|E|\ll|t_1|$ of the Hamiltonian
in Eq.~(\ref{eq:H_bi_0}), the energy and wavefunction are given by
$E_{s\,{\bf k}}\approx s{\hbar^2}q^2/{2m^*}$ where $m^*={\hbar^2
|t_1|}/{2t_0^2}$ and
\begin{equation}
\left|\psi_{s\,{\bf k}}\right>\approx\frac{1}{\sqrt{2}} \left(
\begin{array}{c}
e^{-i\theta_{\bf q}}\\
0\\
0\\
-{\rm sgn}(t_1)\,s\,e^{i\theta_{\bf q}+ik_{\rm z}d}
\end{array}
\right)\,, \label{eq:wfn_bi}
\end{equation}
respectively~\cite{Ando:bilayer}. The phase difference $e^{ik_{\rm
z}d}$ between the two graphene layers in Eq.~(\ref{eq:H_bi_0})
appears here as well. Using Eqs.~(\ref{eq:H_bi_int}),
(\ref{eq:pw_bi}), and~(\ref{eq:wfn_bi}), the transition matrix
element is given for light polarized along the {\it x} direction by
\begin{equation}
M^{x-{\rm pol}}_{s\,{\bf k}}\sim \exp(-i\theta_{\bf q})-{\rm
sgn}(t_1)\,s\,\exp(i\theta_{\bf q}+ik_{\rm z}d)\,. \label{eq:mtx_bi}
\end{equation}
It follows that for $s=+1$ (states above the Dirac point energy),
\begin{equation}
|M^{x-{\rm pol}}_{+1\,{\bf k}}|^2\sim\sin^2(\theta_{\bf q}+k_{\rm
z}d/2) \label{eq:mtx2_bi_1}
\end{equation}
and
\begin{equation}
|M^{x-{\rm pol}}_{+1\,{\bf k}}|^2\sim\cos^2(\theta_{\bf q}+k_{\rm
z}d/2) \label{eq:mtx2_bi_2}
\end{equation}
with $t_1>0$ and $t_1<0$, respectively. The perpendicular component
of the wavevector $k_{\rm z}$ reads~\cite{Shuyun},
\begin{equation}
k_{\rm z}=\sqrt{2m_e(E_{\rm KE}+V_{\rm inner})/\hbar^2-k^2}
\label{eq:k_z}
\end{equation}
where $E_{\rm KE}$ is the kinetic energy of the photo-electron and
$V_{\rm inner}$ the inner potential. Note here that {\bf k} is a
two-dimensional Bloch wavevector, i,\,e.\,, it has no out-of-plane
component. The inner potential has been measured for graphite by
analyzing the energy dispersion at normal emission (i.\,e.\,, ${\bf
k}=0$)~\cite{Shuyun}. According to Eq.~(\ref{eq:k_z}), $k_{\rm
z}d\approx3.9\,\pi$.

Similarly, for light polarized in the {\it y} direction, the
transition matrix element for a double-layer graphene is given by
\begin{equation}
M^{y-{\rm pol}}_{s\,{\bf k}}\sim \exp(-i\theta_{\bf q}) +{\rm
sgn}(t_1)\,s\,\exp(i\theta_{\bf q}+ik_{\rm z}d)\,. \label{eq:mty_bi}
\end{equation}
It follows that for $s=+1$ (states above the Dirac point energy),
\begin{equation}
|M^{y-{\rm pol}}_{+1\,{\bf k}}|^2\sim\cos^2(\theta_{\bf q}+k_{\rm
z}d/2) \label{eq:mty2_bi_1}
\end{equation}
and
\begin{equation}
|M^{y-{\rm pol}}_{+1\,{\bf k}}|^2\sim\sin^2(\theta_{\bf q}+k_{\rm
z}d/2) \label{eq:mty2_bi_2}
\end{equation}
with $t_1>0$ and $t_1<0$, respectively.

In both cases, we can explain the rotation of the intensity maximum
in the photoelectron intensity map around the K point by $\pi/2$
upon the change from X- to Y-polarized light. Comparing
Eqs.~(\ref{eq:mtx2_bi_1}) and~(\ref{eq:mtx2_bi_2}) with
Eqs.~(\ref{eq:mty2_bi_1}) and~(\ref{eq:mty2_bi_2}), irrespective of
the sign of $t_1$, the maxima of the photoemission intensity map of
a double-layer graphene is rotated by $\pi/2$ when the light
polarization is rotated by $\pi/2$, in agreement with experiment
shown in Fig.~\ref{Fig3}. If we assume that $t_1>0$, which is
qualitatively in agreement with experiment shown in
Figs.~\ref{Fig3}(a) and~\ref{Fig3}(b), $I^{x-{\rm pol}}_{+1\,{\bf
k}}\propto|M^{x-{\rm pol}}_{+1\,{\bf k}}|^2\sim\sin^2(\theta_{\bf
q}+k_zd/2)$ and $I^{y-{\rm pol}}_{+1\,{\bf k}}\propto|M^{y-{\rm
pol}}_{+1\,{\bf k}}|^2\sim\cos^2(\theta_{\bf q}+k_zd/2)$ for the
upper band ($s=+1$). Therefore, we have determined that the vertical
inter-layer hopping matrix element $t_1$ between two carbon $p_{\rm
z}$ orbitals sitting on top of each other is positive; we will come
back to this point later.

Although this model can overall account for the experimental data of
double-layer graphene, there is a discrepancy: the measured
photoemission intensity at $E=E_{\rm D}+0.25$~eV along the $+k_x$
direction for the X-polarization geometry is finite, whereas the
theory predicts this value to vanish as shown in the $E_{\rm
D}+0.25$~eV map of the Ideal DG in Fig.~\ref{Fig5}(a). We believe
that this discrepancy arises from the finite size of the light beam
spot ($\sim$80$\times$40 $\mu$m$^2$) which covers not only the
double-layer graphene portion but also some single-layer graphene,
as discussed in Fig.~\ref{Fig1}. In fact, double-layer graphene
samples inevitably contain a finite amount of single-layer
graphene~\cite{David,Ohta2}. The fraction of single-layer coverage
can be obtained by LEEM measurements~\cite{David,Ohta2}: from this
analysis shown in Fig.~\ref{Fig1}, we find that the double-layer
graphene sample used here contains $\sim$74\% double-layer and
$\sim$22\% single-layer graphene. When the theoretical photoelectron
intensity maps of single- and double-layer graphene are
correspondingly weighted and averaged, the results denoted by
SG$+$DG in Figs.~\ref{Fig5}(a) and~\ref{Fig5}(b) are in excellent
agreement with the experimental data shown in Figs.~\ref{Fig3}(a)
and~\ref{Fig3}(b). This is even more clear from the angular
dependence of the photoelectron intensity maps shown in the red
solid lines in Fig.~\ref{Fig5}(c) and~\ref{Fig5}(d). Note that the
presence of single-layer graphene does not affect the results for
the Y-polarization, which is obvious when we compare the red solid
and green dashed lines in Fig.~\ref{Fig5}(d), because both the
intensity maxima of single- and double-layer graphene occur near
$\theta$ = $\pi$.

\section{Discussion}
\subsection{Berry's phase}
We have shown that, when the light polarization is rotated by
$\pi/2$, the maximum intensity position in $I_{\bf k}$ in the
$k_x$-$k_y$ plane of single- and double-layer graphene is rotated by
$\pi/n$, where $n=1$ and $n=2$ for single- and double-layer
graphene, respectively. The physical meaning of these rotations,
whose origins rests on the phase factor $\exp\left(\pm i\,
n\,\theta_{\bf q}/2\right)$ of the sublattice amplitude of the
wavefunctions~\cite{Antonio}, becomes clear upon a complete
circulation of {\bf q} around the Dirac point K ($\theta_{\bf
q}\,\rightarrow\,\theta_{\bf q}\,+\,2\pi$), which directly gives a
Berry's phase $\beta=n\,\Delta\theta_{\bf
q}/2=n\,\pi$~\cite{Antonio}. Recently the Berry's phase
interpretation of $n$~\cite{McCann,Zhang,Novoselov} has been given
by a different interpretation in terms of pseudospin winding
number~\cite{Parkcond}.

The fact that $n\,\theta_{\bf q}/2$ enters in $I_{\bf k}$ in the
form of either $\sin^2\left(n\,\theta_{\bf q}/2+\varphi\right)$ or
$\cos^2\left(n\,\theta_{\bf q}/2+\varphi\right)$ with some constant
$\varphi$, demonstrates that the matrix elements directly contain
information on the Berry's phase. The rotation of light polarization
gives an additional phase $\pi/2$ to the phase factor, i.\,e.\,,
$\exp\left(\pm i\,n\,\theta_{\bf
q}/2\right)\,\rightarrow\,\exp\left(\pm i\,n\,\theta_{\bf
q}/2\,+\,i\,\pi/2\right)$. Thus, the photoelectron intensity $I_{\bf
k}$ is modified accordingly from $\sin^2\left(n\,\theta_{\bf
q}/2+\varphi\right)\rightarrow \cos^2\left(n\,\theta_{\bf
q}/2+\varphi\right)$ or $\cos^2\left(n\,\theta_{\bf
q}/2+\varphi\right)\rightarrow\sin^2\left(n\,\theta_{\bf
q}/2+\varphi\right)$ upon the rotation of light polarization by
$\pi/2$, i.\,e.\,, the $\pi/n$ rotation of $I_{\bf k}$. This
prediction is exactly realized in our experimental results.

The power of our method is that it can be extended in a
straightforward way to other materials with Berry's phase
$\beta=n\,\pi$ (not necessarily~$\pi$ or~$2\pi$). In this case, the
photoemission intensities for X- and Y-polarization geometries are
given by $I^{x-{\rm pol}}_{\bf k}\propto\sin^2\left(n\,\theta_{\bf
q}/2+\varphi\right)$ and $I^{y-{\rm pol}}_{\bf
k}\propto\cos^2\left(n\,\theta_{\bf q}/2+\varphi\right)$, where
$\varphi$ is a system-dependent constant. The important feature is
that the rotation of light polarization by $\pi/2$ results in a
rotation of intensity maxima by $\pi/n$ for a system with
$\beta=n\,\pi$ regardless of the constant $\varphi$. Thus, we have
demonstrated here that the Berry's phase can directly be measured
from polarization-dependent ARPES.

Unlike methods based on magneto-transport
experiments~\cite{Zhang,Novoselov}, our new method has three
important advantages. (i) The Berry's phase of a specific electronic
band can be measured, because ARPES has the angle-resolving power
and also because one can set up a tight-binding Hamiltonian
focussing on only the electronic states of interest: those two
results can directly be compared with each other. (ii) Due to the
angle-resolving power, the measured result is free from valley
degeneracy for the case of graphene. (iii) Our method does not need
electric gating, which is essential for the transport measurements.

\subsection{The sign of inter-orbital hopping integrals}
\begin{figure}
  \begin{center}
  \includegraphics[width=0.9\columnwidth]{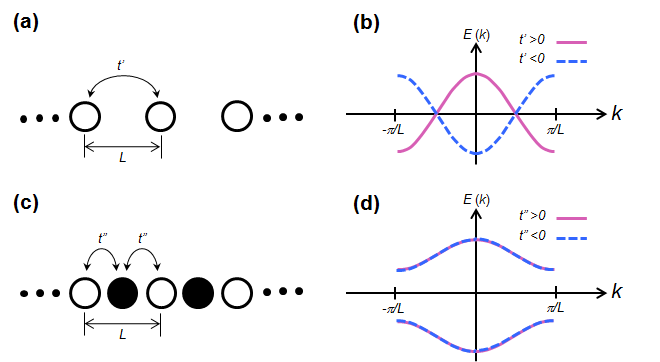}
  \end{center}
  \caption{
(a) Schematic of a one-dimensional crystal having {\it s}-type
orbital per unit cell. The nearest neighbor hopping integral is
$t^{\prime}$. (b) Calculated electron energy band structure of the
system depicted in (a) with different choices for the sign of
$t^{\prime}$. (c) Schematic of a one-dimensional crystal having two
{\it nonequivalent} {\it s}-type orbitals per unit cell (i.\,e.\,,
having different on-site energies). We assume that all the distances
between the centers of the nearest neighbor orbitals are the same
and hence the corresponding hopping integrals, denoted by
$t^{\prime\prime}$. (d) Calculated electronic band structure of the
system depicted in {\bf c} with different choices for the sign of
$t^{\prime\prime}$.}
  \label{Fig6}
  \end{figure}

Another important finding of our study is that we can directly
extract, for the first time, the sign and the absolute magnitude of
the inter-orbital hopping integrals (IOHIs) between non-equivalent
localized orbitals of a tight-binding Hamiltonian from experiment.
Until now, in fact, the sign determination of an IOHI has resorted
no to any experimental method, but to {\it ab initio} calculations,
e.\,g.\,, using maximally localized Wannier
functions~\cite{Marzari}. In order to understand an ambiguity
related with the experimental sign determination, we take the
simplest one-dimensional example, and extend the discussion to a
more complicated tight-binding model of graphitic systems than the
one described previously. We consider simple tight-binding models
having {\it s}-like localized states, the values of which are all
positive in real space (we can arbitrarily set this gauge without
losing generality.) If there is only one localized orbital per unit
cell in a one-dimensional tight-binding model as drawn in
Fig.~\ref{Fig6}(a), the energy bandstructure varies with the sign of
the IOHI $t^{\prime}$ as shown in Fig.~\ref{Fig6}(b). Hence, the
sign of IOHIs between ``equivalent'' orbitals can always be
trivially determined~\cite{Hasan}.

However, if we consider the case where there are two non-equivalent
localized orbitals $\phi_{s}$ and $\phi'_{s}$ whose value in real
space is positive and if we denote the IOHI between the nearest
neighboring orbitals by $t^{\prime\prime}$ as drawn in
Fig.~\ref{Fig6}(c), the actual band structure is invariant when we
change the sign of $t^{\prime\prime}$ as shown in
Fig.~\ref{Fig6}(d). Therefore, even when the actual electronic band
structure is empirically determined, the sign of $t^{\prime\prime}$
cannot be determined. In general, an empirical tight-binding model
with more than one orbital per unit cell has this sign ambiguity
problem for IOHIs between ``non-equivalent'' orbitals, thus
preventing an experimental measurement of IOHI from just the energy
band dispersions.

%%%%%%%%%%%%%%%%%%%%%%%%%%%%%%%%%%%%%%%%%%%%%%%%%%%%%%%%%%%%%%%%%%%%55
  \begin{figure*}
  \begin{center}
  \includegraphics[width=1.8\columnwidth]{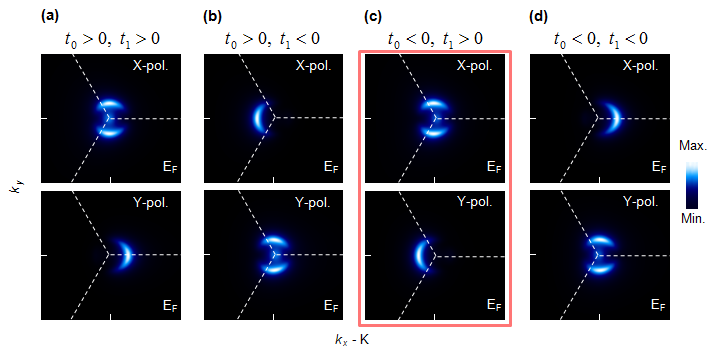}
  \end{center}
  \caption{(a-d) Calculated intensity maps for four different
choices of the signs of nearest-neighbor in-plane and vertical
inter-layer hopping integrals, $t_0$ and $t_1$, respectively. Case
(c) agrees with the experimental results.}
  \label{Fig7}
  \end{figure*}
%%%%%%%%%%%%%%%%%%%%%%%%%%%%%%%%%%%%%%%%%%%%%%%%%%%%%%%%%%%%%%%%%%%%

We could understand this degeneracy as follows. If we denote the
Bloch sums of the two localized orbitals by $\phi_1({\bf k})$ and
$\phi_2({\bf k})$, the tight-binding Hamiltonian $H$ using this
basis set reads
\begin{equation}
H({\bf k})=\left(
\begin{array}{rr}
\left<\phi_1({\bf k})\right| H \left|\phi_1({\bf k})\right> &
\left<\phi_1({\bf k})\right| H \left|\phi_2({\bf k})\right> \\
\left<\phi_2({\bf k})\right| H \left|\phi_1({\bf k})\right> &
\left<\phi_2({\bf k})\right| H \left|\phi_2({\bf k})\right> \\
\end{array}
\right)\,. \label{eq:H_phi12}
\end{equation}
Note here that the matrix elements $\left<\phi_i({\bf
k})\right|H\left|\phi_j({\bf k})\right>$, in which $i,j\in\{1,2\}$,
involve not only the on-site or nearest-neighbor hopping processes
but also all the other possible hopping processes. Now, it is
obvious that the following Hamiltonian $H'({\bf k})$ has exactly the
same eigenvalues as $H({\bf k})$:
\begin{equation}
H'({\bf k})=\left(
\begin{array}{rr}
\left<\phi_1({\bf k})\right| H \left|\phi_1({\bf k})\right> &
-\left<\phi_1({\bf k})\right| H \left|\phi_2({\bf k})\right> \\
-\left<\phi_2({\bf k})\right| H \left|\phi_1({\bf k})\right> &
\left<\phi_2({\bf k})\right| H \left|\phi_2({\bf k})\right> \\
\end{array}
\right)\,. \label{eq:Hp_phi12}
\end{equation}
What we have done by going from $H({\bf k})$ to $H'({\bf k})$ is to
change the sign of the IOHI between the two non-equivalent localized
orbitals. In fact, the two matrices $H({\bf k})$ and $H'({\bf k})$
are related by a unitary transform, which does not change the
eigenvalues of a matrix, $H'({\bf k})=U^\dagger H({\bf k})U$ with
\begin{equation}
U=\left(
\begin{array}{rr}
1 & 0 \\
0 & -1
\end{array}
\right)\,.
\end{equation}
On the other hand, if we change the signs of the diagonal matrix
elements $\left<\phi_i({\bf k})\right|H\left|\phi_i({\bf
k})\right>$, in which $i\in\{1,2\}$, we get a different eigenvalue
spectrum. Thus, there is no ambiguity in the sign of the IOHI
between equivalent orbitals. This simple example illustrates that
one cannot determine the signs of the IOHIs between non-equivalent
orbitals just by looking at the measured electron energy
bandstructure.

The tight-binding model for double-layer graphene that we used in
our calculations is based on the $p_z$ orbitals of carbon atoms with
two parameters: $t_0$ and $t_1$ for the nearest-neighbor in-plane
and the vertical inter-layer hopping integrals, respectively. The
parameters $t_0$ and $t_1$ correspond to $-\gamma_0'$ and
$\gamma_1'$ in the well-known Slonczewski-Weiss-McClure model,
respectively~\cite{SWMc1,SWMc2}, and we have used $\vert
t_0\vert=3.16$~eV and $\vert t_1\vert=0.39$~eV~\cite{Gruneis}. The
photoemission intensity map in the $k_x$-$k_y$ plane is strongly
dependent on the signs of both $t_0$ and $t_1$ as shown in
Fig.~\ref{Fig7}. Because the four different choices of the signs
produce exactly the same electronic band structure, it has not been
known from experiments which choice of the signs of the IOHIs is
physically correct, although the absolute values have been
experimentally estimated~\cite{Dresselhaus1,Dresselhaus2}.

This sign-ambiguity problem still exists even when we include more
complicated hopping processes in the model, especially the
non-vertical inter-layer hopping integrals $\gamma_3'$ and
$\gamma_4'$, there still exist unitary transforms that leave the
energy eigenvalues unchanged. In a tight-binding model having four
hopping integrals ($\gamma_0'$, $\gamma_1'$, $\gamma_3'$, and
$\gamma_4'$), the following four different sets of parameters give
exactly the same electron energy bandstructure: ($\gamma_0'$,
$\gamma_1'$, $\gamma_3'$, $\gamma_4'$), ($\gamma_0'$, $-\gamma_1'$,
$-\gamma_3'$, $-\gamma_4'$), ($-\gamma_0'$, $\gamma_1'$,
$\gamma_3'$, $-\gamma_4'$), and ($-\gamma_0'$, $-\gamma_1'$,
$-\gamma_3'$, $\gamma_4'$), assuming that the first set is composed
of the values currently accepted and used when the SWMc model is
considered. In principle, there should be eight different sets of
parameters giving the same energy bandstructure; however, from our
knowledge that the nearest-neighbor intralayer hopping integrals in
different graphene layers are the same, we have reduced the number
of candidates to four. This is also the reason why we considered
only the four cases in Fig.~\ref{Fig7}.

%Since we have shown that the choice of $t_0<0$ and $t_1>0$,
%i.\,e.\,, $\gamma'_0>0$ and $\gamma'_1>0$ in Fig.~\ref{Fig7}(c),
%reproduces the experimental results shown in Figs.~\ref{Fig3}(a)
%and~\ref{Fig3}(b), we can resultantly determine $\gamma'_3>0$ and
%$\gamma'_4>0$.
We have shown that the choice of $t_0<0$ and $t_1>0$, i.\,e.\,,
$\gamma'_0>0$ and $\gamma'_1>0$ in Fig.~\ref{Fig7}(c), reproduces
the experimental results shown in Figs.~\ref{Fig3}(a)
and~\ref{Fig3}(b), thus experimentally determined the signs of these
IOHIs. The signs of the IOHIs used for graphite in the conventional
model~\cite{SWMc1,SWMc2} is indeed correct. Previous theoretical
study has also tried to determine the signs~\cite{Eli}, but
inappropriate theoretical approaches (as previously mentioned) and
the lack of full polarization-dependent experimental data have led
to incorrect speculations, which do not agree with the conventional
model~\cite{SWMc1,SWMc2} as well as our results. Our method can
generally be used to determine the sign of the hopping integrals in
complicated materials such as cuprates as well as in simple
materials such as gallium arsenide and one-dimensional crystals
having two atoms per unit cell.

\section{Summary}
We have shown that ARPES can be used as a powerful tool to directly
measure quantum phases such as the Berry's phase of a specific
electronic band in graphene with advantages compared to the
interference type of measurements~\cite{Swerner,Tomita,Bitter} which
do not give any information on the band-specific Berry's phase, and
the sign of the hopping integral between non-equivalent orbitals,
never measured for any material before. The experimental and
theoretical procedures developed here can be applied in studying the
electronic, transport, and quantum interference properties of a huge
variety of materials.

%{\bf {\it Note added}: During the submission process, we became
%aware that the understanding of Berry's phase in
%graphene~\cite{McCann} used in the current manuscript, along with
%that commonly used in the widespread community, may need adjustment.
%It has recently been found that experiments including the present
%work as well as previous works~\cite{Novoselov} are sensitive to a
%material parameter more accurately described as the pseudo spin
%winding number instead of the Berry's phase.  More work will be
%required to distinguish these ideas.}

\acknowledgments We gratefully acknowledge D.-H. Lee, J. Graf, C. M.
Jozwiak, S. Y. Zhou and H. Zhai for helpful discussions. This work
was supported by the Director, Office of Science, Office of Basic
Energy Sciences, Materials Sciences and Engineering Division, of the
U.S. Department of Energy under Contract No. DE-AC02-05CH11231.

\end{document}